\documentclass[lettersize,journal]{IEEEtran}
\usepackage{amsmath,amsfonts}
\usepackage{algorithmic}
\usepackage{algorithm}
\usepackage{array}
\usepackage[caption=false,font=normalsize,labelfont=sf,textfont=sf]{subfig}
\usepackage{textcomp}
\usepackage{stfloats}
\usepackage{url}
\usepackage{verbatim}
\usepackage{graphicx}
\usepackage{cite}
\usepackage{multirow}
\usepackage{booktabs}
\usepackage{xcolor} 
\usepackage[table]{xcolor}
\definecolor{myblue}{HTML}{0057FF}

\begin{document}

\title{ComGPT: Detecting Local Community Structure with Large Language Models}

\author{Li Ni, Haowen Shen, Lin Mu, Yiwen Zhang*, and Wenjian Luo, Senior Member, IEEE
\IEEEcompsocitemizethanks{
\IEEEcompsocthanksitem 
Li Ni is with Key Laboratory of Intelligent Computing \& Signal Processing, Ministry of Education and School of Computer Science and Technology, Anhui University, Hefei, Anhui, 230601, China. Li Ni is also with Guangdong Provincial Key Laboratory of Novel Security Intelligence Technologies, Shenzhen, 518055, China.

 Haowen Shen, Lin Mu and Yiwen Zhang are with the School of Computer Science and Technology, Anhui University, Hefei, Anhui, 230601, China. 

Wenjian~Luo is with Guangdong Provincial Key Laboratory of Novel Security Intelligence Technologies, Institute of Cyberspace Security, School of Computer Science and Technology, Harbin Institute of Technology, Shenzhen 518055, Guangdong, China.

Email: nili@ahu.edu.cn, e23201075@stu.ahu.edu.cn, mulin@ahu.edu.cn, zhangyiwen@ahu.edu.cn, luowenjian@hit.edu.cn. (Corresponding author: Yiwen Zhang)}}

\maketitle

\begin{abstract}
Large Language Models (LLMs), like GPT-3.5-turbo, have demonstrated the ability to understand graph structures and have achieved excellent performance in various graph reasoning tasks, such as node classification. Despite their strong abilities in graph reasoning tasks, they lack specific domain knowledge and have a weaker understanding of community-related graph information, which hinders their capabilities in the community detection task. Moreover, local community detection algorithms based on seed expansion, referred to as seed expansion algorithms, often face  several shortcomings, including the seed-dependent problem, community diffusion, and free rider effect.
To use LLMs to overcome the above shortcomings, we explore a GPT-guided seed expansion algorithm named  ComGPT. ComGPT iteratively selects potential nodes by  local modularity  from the detected community's neighbors, and subsequently employs LLMs to choose the node from these selected potential nodes to join the detected community. 
To improve LLMs' understanding of community-related graph information, we propose ComIncident, a graph encoding method that incorporates community knowledge and  is designed for the community detection task.
Additionally, we design the  Node Selection Guide (NSG) prompt  to enhance LLMs' understanding of community characteristics. Experimental results demonstrate that ComGPT outperforms the baselines, thereby confirming the effectiveness of the ComIncident and the NSG prompt.
\end{abstract}


\section{Introduction}
\IEEEPARstart{I}{dentifying} community structures from networks has wide applications, such as identifying protein complexes \cite{yin2017local}. Local community detection has gained attention due to its ability to quickly identify the community containing a specific node \cite{10164164,LSADEN}. Various methods are designed for detecting local community, including seed expansion algorithms \cite{M},  pagerank based techniques \cite{4031383}, flow-propagation based techniques \cite{DBLP:conf/sdm/VeldtKG19,10.5555/3495724.3496146}, and non-negative matrix factorization algorithms \cite{DBLP:journals/datamine/WangLWZD11}. Among these methods, the seed expansion algorithms have garnered widespread acclaim for their versatility and ease of implementation. It typically enlarges the community by selecting a node with the high scoring function value at each step \cite{moradi2014local}. 


\begin{figure}[!t] 
    \centering
 \includegraphics[width=0.8\linewidth]{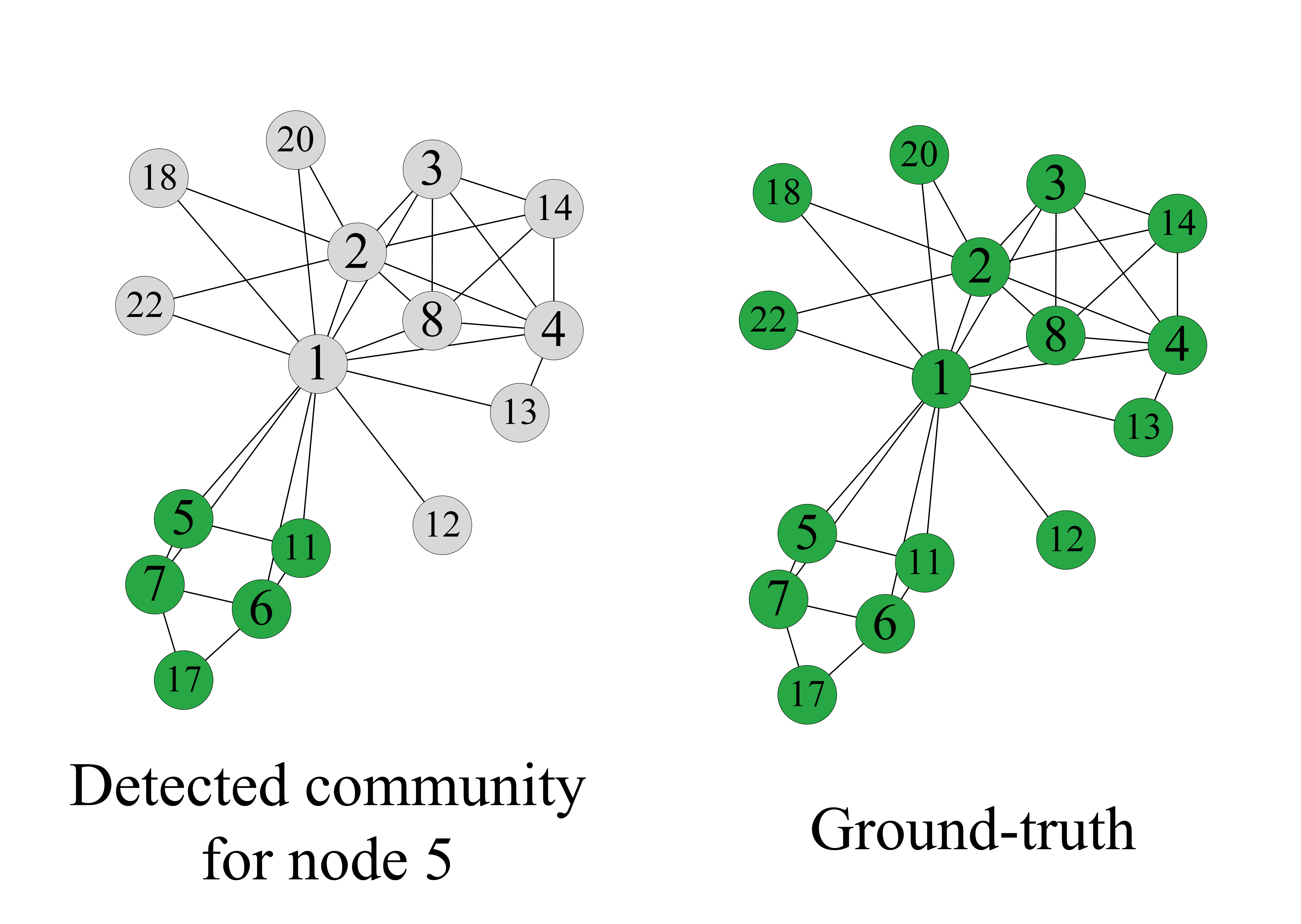} 
    \caption{Seed-dependent problem. The community detected by M method \cite{M} with starting nodes 5 is \{5, 11, 6, 7, 17\}. This community misses many nodes. Nodes in the same community are marked with the same color.}
    \label{Seed_dependent_problem}
\end{figure}

\begin{figure}[!t] 
    \centering
 \includegraphics[width=0.8\linewidth]{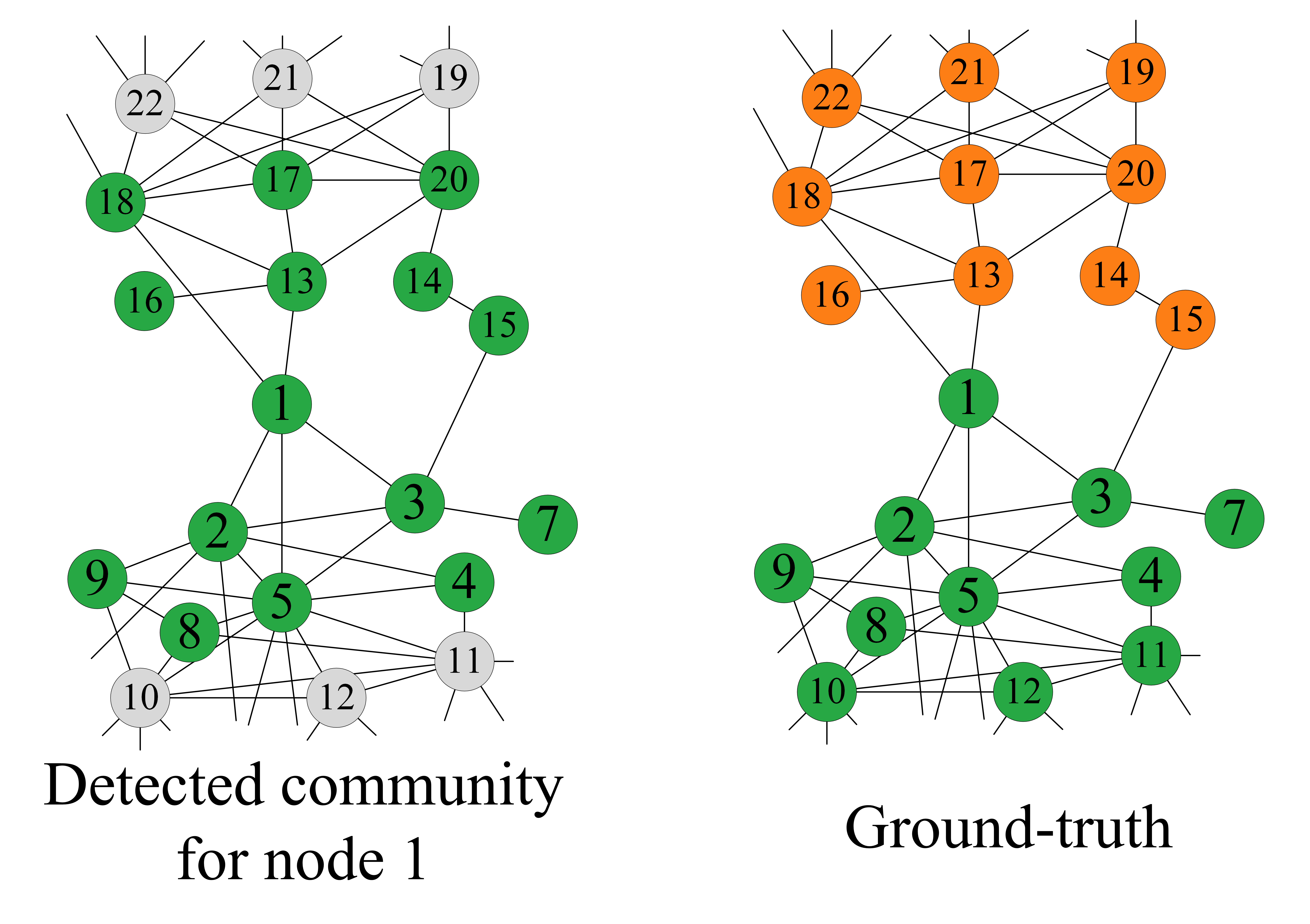} 
    \caption{Community diffusion. The community detected by the M method  \cite{M} with node 1 as the starting seed node is \{1, 2, 3, 4, 7, 8, 9, 13, 14, 15, 16, 17, 18, 20\}, which incorrectly includes nodes from two different communities. Nodes in the same community are marked with the same color.}
    \label{community_diffusion}
\end{figure}

\begin{figure}[!t] 
    \centering
 \includegraphics[width=0.8\linewidth]{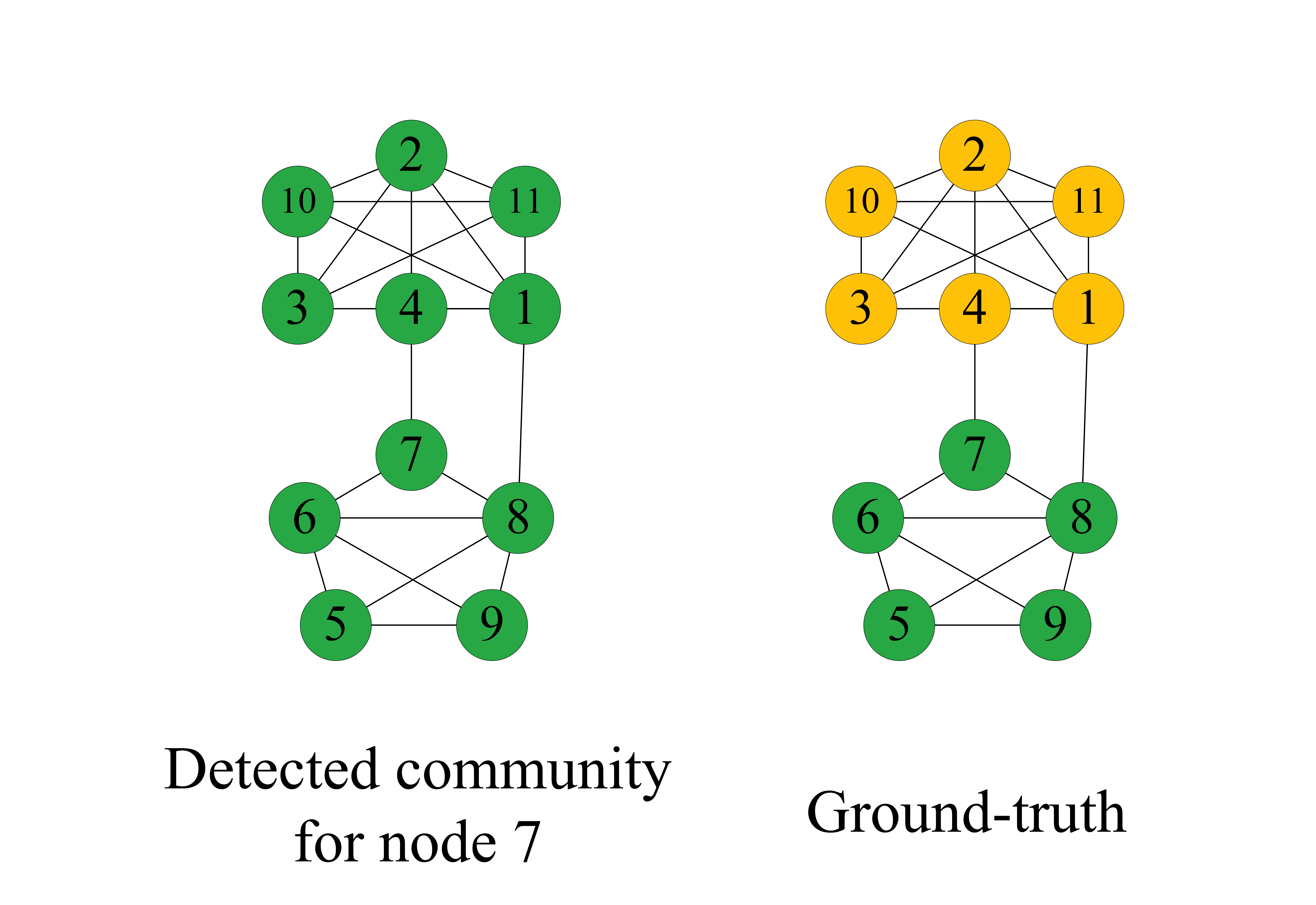} 
    \caption{Free rider effect. The target community consists of nodes 5, 6, 7, 8, and 9. The detected community not only contains nodes in the target community \{5, 6, 7, 8, 9\} but also erroneously includes all nodes in the unrelated  community \{1, 2, 3, 4, 10, 11\}. Nodes in the same community are marked with the same color.}
    \label{free_rider_effect}
\end{figure}




However, seed expansion algorithms often face the seed-dependent problem \cite{10.1145/2623330.2623621}, community diffusion, and free rider effect \cite{10.1145/1835804.1835923,10.1145/1321440.1321526,10.1145/2487575.248764}, which are  shown in Figure \ref{Seed_dependent_problem}, \ref{community_diffusion}, and \ref{free_rider_effect}, respectively. The seed-dependent problem refers to the quality of the community detected being highly dependent on the initial seed nodes, as illustrated by node 5 in Figure \ref{Seed_dependent_problem}. The community diffusion problem means that the detected community contains nodes from different communities,  as illustrated by node 1 in Figure \ref{community_diffusion}. The free rider effect refers to mistakenly adding unrelated communities to enhance the scoring function value of a detected community, as illustrated by node 7 in Figure \ref{free_rider_effect}. To address these problems, researchers developed different strategies such as identifying core node sets \cite{LCDMD}, finding the nearest nodes with greater centrality \cite{LCDNN},  alternating strategy of strong fusion and weak fusion \cite{ASFWF}, and so on. 

The above studies rely on predefined heuristic rules, which may fail to adapt to diverse structures across networks. 
Recently, Large Language Models (LLMs) have emerged as powerful reasoning engines \cite{ijcai2024-1,ijcai2024-2,ijcai2024-3,adhikari2020learning,ammanabrolu2021learning,wang2024can}.
Unlike these heuristic-based approaches, LLMs do not rely on a fixed set of rules and can  be prompted to reason about diverse network structures.
Therefore, we explore a GPT-guided seed expansion approach based on LLMs, called ComGPT.
ComGPT leverages LLMs to guide the detection of a local community.
The one-question-one-answer modality of LLMs aligns well with the incremental nature of seed expansion algorithms, making it naturally suitable for this task. This approach  bridges the current gap in LLMs applicability to the community detection domain.



Specifically, we leverage LLMs to select nodes to join the community instead of the scoring function used in seed expansion algorithms.
However, the application of LLMs to select nodes faces the following challenges: 
\begin{enumerate}
    \item Lack of graph encoding methods for providing LLMs with sufficient community-related graph information.
To address this, we propose ComIncident (Community-Enriched Incident), a variant of the graph encoding method Incident \cite{fatemi2023talk}, which incorporates community information.

   \item Lack of domain knowledge for  LLMs in dealing with local community detection. It limits the capabilities of LLMs. To address this, we design prompts to assist LLMs in understanding community definitions and how to select nodes. 
   
  \item Token count constraints.  To address this,  LLMs receive only a partial network as input, not the entire network, to minimize costs by reducing the number of tokens processed.
\end{enumerate}

Our contributions are summarized as follows:

\begin{enumerate}
    \item 
    We propose a GPT-guided seed expansion algorithm called ComGPT, which iteratively uses GPT-3.5-turbo to add nodes to the community. It includes four main steps: potential node identification, graph encoding, node selection by GPT-3.5-turbo, and node supplementation by GPT-3.5-turbo. To the best of our knowledge, this is the first work to apply LLMs to the domain of community detection.
    
    \item 
    We design an  graph encoding method, named ComIncident, and NSG (Node Selection Guide) prompt. ComIncident  is the first graph encoding method designed for the community detection task, which enhances LLMs to understand the current state of the detected community. The NSG prompt, which integrates community features, facilitates LLMs's comprehension of domain knowledge. 
    
    \item The results on real datasets indicate that ComGPT outperforms baselines. The ablation study experiments confirm that LLMs play a crucial role in the ComGPT's performance. Furthermore, prompt comparison experiments reveal that the designed prompt effectively aids LLMs in understanding the graph structure and community situation.

\end{enumerate}

The rest of the paper is organized as follows. Section \ref{sec:Approach} presents the proposed algorithm. Section \ref{sec:Experiments} reports experimental results. Section \ref{sec:Related work} summarizes the related works. Section \ref{sec:Conclusion} concludes the paper.


\section{Approach}
\label{sec:Approach}

\begin{figure*}[!t] 
    \centering
\includegraphics[width=0.79\linewidth]{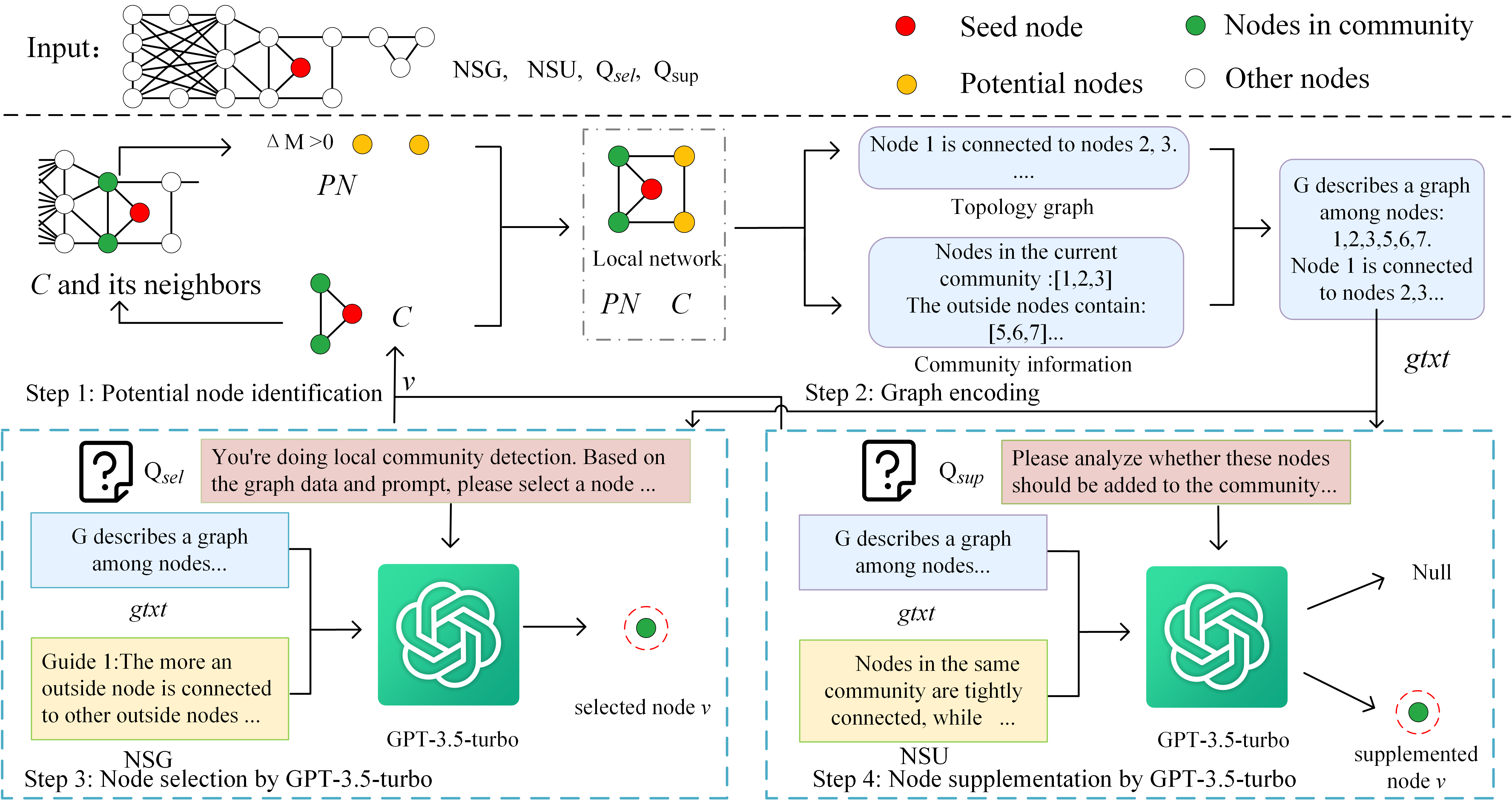} 
    \caption{The diagram of ComGPT. It includes four main steps: 1) Potential node identification aims to pinpoint potential nodes from the neighbors of the current community. 2) Graph encoding involves transforming the graph structure into a textual format interpretable by GPT-3.5-turbo. 3) Node selection by GPT-3.5-turbo is that GPT-3.5-turbo selects the node that optimally integrates into the community from potential nodes. 4) Node supplementation by GPT-3.5-turbo involves GPT-3.5-turbo selecting which nodes to add to the community when no potential nodes remain.}
    \label{fig:ComGPT}
\end{figure*}

We introduce a GPT-guided seed expansion algorithm named ComGPT.
This section begins with an overview of ComGPT, followed by detailed descriptions of each step of ComGPT. Table \ref{tab:notations} lists the important notations used in the paper.

\subsection{Overview}


\begin{table}[!t]
\centering
\caption{Important notations}
\begin{tabular}{p{0.17\columnwidth}<{\centering}| p{0.7\columnwidth}} 
\toprule
Notation & Definition\\ 
\midrule
$G$ & network\\
$C$ & community\\
$N$ & neighbor node of $C$\\
$v_s$ & seed node\\
$PN$ & potential node \\
$cands$ &  communities each time $PN$ is empty\\
$gtxt$ & graph text obtained by the graph encoding step \\
$NSG$ & node selection guide prompt\\
$NSU$ & node supplementation prompt \\
$Q_{sel}$ & node selection question\\
$Q_{sup}$ & node supplementation question \\
\bottomrule
\end{tabular}
\label{tab:notations}
\end{table}

As shown in Figure \ref{fig:ComGPT},  ComGPT contains four main steps: potential node identification, graph encoding, node selection by GPT-3.5-turbo, and node supplementation by GPT-3.5-turbo.
The idea of ComGPT uses  potential node identification step (Section \ref{sec:Potential nodes identification}) to select  potential nodes from the neighbors of the current community.  Then, it uses node selection by GPT-3.5-turbo step (Section \ref{sec:Node selection by GPT-3.5-turbo}) and  node supplementation by GPT-3.5-turbo step (Section \ref{sec:Node supplementation by GPT-3.5-turbo}) selects nodes to join the community. 
Since GPT-3.5-turbo cannot directly process graph structures and community structures, the graph encoding step (Section \ref{sec:Graph encoding}) is used to encode the graph structure and community structure into text that GPT-3.5-turbo can understand.

Algorithm \ref{alg:ComGPT} shows the overall process of ComGPT. Initially,  community $C$ contains seed node $v_s$ and $cands$  is empty (line \ref{alg:ComGPT_Init}), where $cands$ stores the communities each time $PN$ is empty. First, ComGPT selects potential nodes $PN$  with the step of potential node identification (line \ref{alg:ComGPT_fpn1}), used for subsequent graph encoding.
In order for GPT-3.5-turbo to understand the graph structure and community information, the local network, i.e., the subgraph consisting of the nodes in both current community $C$ and $PN$ as well as edges between these nodes, is converted into text, i.e., $gtxt$ (line \ref{alg:ComGPT_graphencoder}). Utilizing $gtxt$, GPT-3.5-turbo selects  node  $v$ from potential nodes $PN$ and adds it to the current community $C$ (lines \ref{alg:ComGPT_nodeselect}-\ref{alg:ComGPT_addv}). 
The above steps are repeated until $PN$ is empty. Then the community $C$  is added to the  $cands$ (line \ref{alg:ComGPT_addc1}). $PN$ being empty does not necessarily imply that the neighboring nodes of the community are weakly connected to it.  Therefore, ComGPT continues to identify nodes through node supplementation (lines \ref{alg:N}-\ref{alg:ComGPT_nodesupplement}). 
If GPT-3.5-turbo identifies suitable node $v$,  ComGPT adds $v$ to $C$ and continues to expand current community $C$ (line \ref{alg:ComGPT_supp}). 
Otherwise,  GPT-3.5-turbo returns ``null" and ComGPT stops community expansion (line \ref{alg:ComGPT_break}).
After the above process is completed, ComGPT selects the community with the highest local modularity $M$ calculated by formula (\ref{eql:M})  from $cands$ as the final community (line \ref{alg:ComGPT_getmaxm}).
During the  expansion process, if no new node is selected in potential nodes identification step (Section \ref{sec:Potential nodes identification}), i.e., $PN$ is empty (line \ref{alg:ComGPT_fpn1}), ComGPT jumps out of the inner while loop. Thereafter, if the node selected in the node supplementation by GPT-3.5-turbo is null (line \ref{v:null}), ComGPT terminates.

\begin{algorithm}[!t]
\caption{ComGPT}
\label{alg:ComGPT}
\textbf{Input}: Network $G$, Seed node $v_{s}$, Prompts NSG and NSU for node selection and supplementation, respectively, Questions Q$_{sel}$ and Q$_{sup}$ for node selection and supplementation, respectively\\
\textbf{Output}: Community $C$
\begin{algorithmic}[1] 
\STATE $C  \leftarrow{\{ v_{s}\}}$, $cands \leftarrow \emptyset$ 
\label{alg:ComGPT_Init}
\WHILE{True} 
\STATE $PN \leftarrow $ potential nodes identification with $C$ 
// Section \ref{sec:Potential nodes identification}
\label{alg:ComGPT_fpn1}
\WHILE{$PN \neq \emptyset$}  \label{PNempty}
\STATE $gtxt \leftarrow$ graph encoding with ($C\cup PN$, $C$, $PN$)   
// Section \ref{sec:Graph encoding}
\label{alg:ComGPT_graphencoder}
\STATE $v \leftarrow $ node selection by GPT-3.5-turbo with ($gtxt$, NSG, Q$_{sel}$ ) // Section \ref{sec:Node selection by GPT-3.5-turbo}
\label{alg:ComGPT_nodeselect}
\STATE add $v$ to $C$
\label{alg:ComGPT_addv}
\label{alg:ComGPT_fpn2}
\ENDWHILE
\STATE add $C$ to $cands$
\label{alg:ComGPT_addc1}
\STATE  $N \leftarrow $   first and second-order neighbors of  $C$ \label{alg:N}
\STATE  $PN \leftarrow $    nodes selected by modularity  $M$ 
\STATE $gtxt \leftarrow$ graph encoding with ($C\cup N$, $C$, $PN$) // Section \ref{sec:Graph encoding} \label{alg:gtxt}
\STATE $ v \leftarrow$ node supplementation by GPT-3.5-turbo with ($gtxt$, NSU, Q$_{sup}$) // Section \ref{sec:Node supplementation by GPT-3.5-turbo}
\label{alg:ComGPT_nodesupplement}
\IF{$v$ is not null} \label{v:null}
\STATE add $v$ to $C$
\label{alg:ComGPT_supp}
\label{alg:ComGPT_fpn3}
\ELSE
\label{alg:ComGPT_addc2}
\STATE break
\label{alg:ComGPT_break}
\ENDIF
\ENDWHILE
\STATE $C \leftarrow$ community with highest $ M$ in $cands$ 
\label{alg:ComGPT_getmaxm}
\RETURN $C$ 
\end{algorithmic}
\end{algorithm}

\subsection{Potential nodes identification} 
\label{sec:Potential nodes identification}

To mitigate the  cost associated with processing nodes through GPT-3.5-turbo, ComGPT identifies potential nodes $PN$ that may belong to the current community and provides $PN$ to GPT-3.5-turbo for selection.
Different scoring functions are designed in the seed expansion algorithms to select nodes \cite{moradi2014local}, which can be used to choose potential nodes,
such as  modularity $M$ \cite{M}, $R$ \cite{clauset2005finding}, and $Q_{lcd}$ \cite{ASFWF}. 
ComGPT uses local modularity $M_C$ \cite{M} to select potential nodes. $M_C$ is calculated as,
\begin{equation} \label{eql:M}
M_C=\frac{e_{ic}}{e_{oc}},
\end{equation}
where $e_{ic}$ represents the number of internal edges in community $C$ and $e_{oc}$ represents the number of external edges of community $C$. For each first-order neighbor node $v$ of  community $C$, if node $v$  satisfies $\Delta M = M_{C \cup \{v\}}- M_{C}> 0$, then it is considered a potential node.
To further reduce the number of nodes processed by GPT-3.5-turbo, $k$ nodes with the largest $M$ are selected from the above  potential nodes to form a potential node's set $PN$.

\subsection{Graph encoding}
\label{sec:Graph encoding}

\begin{figure}[!t] 
    \centering
 \includegraphics[width=1\linewidth]{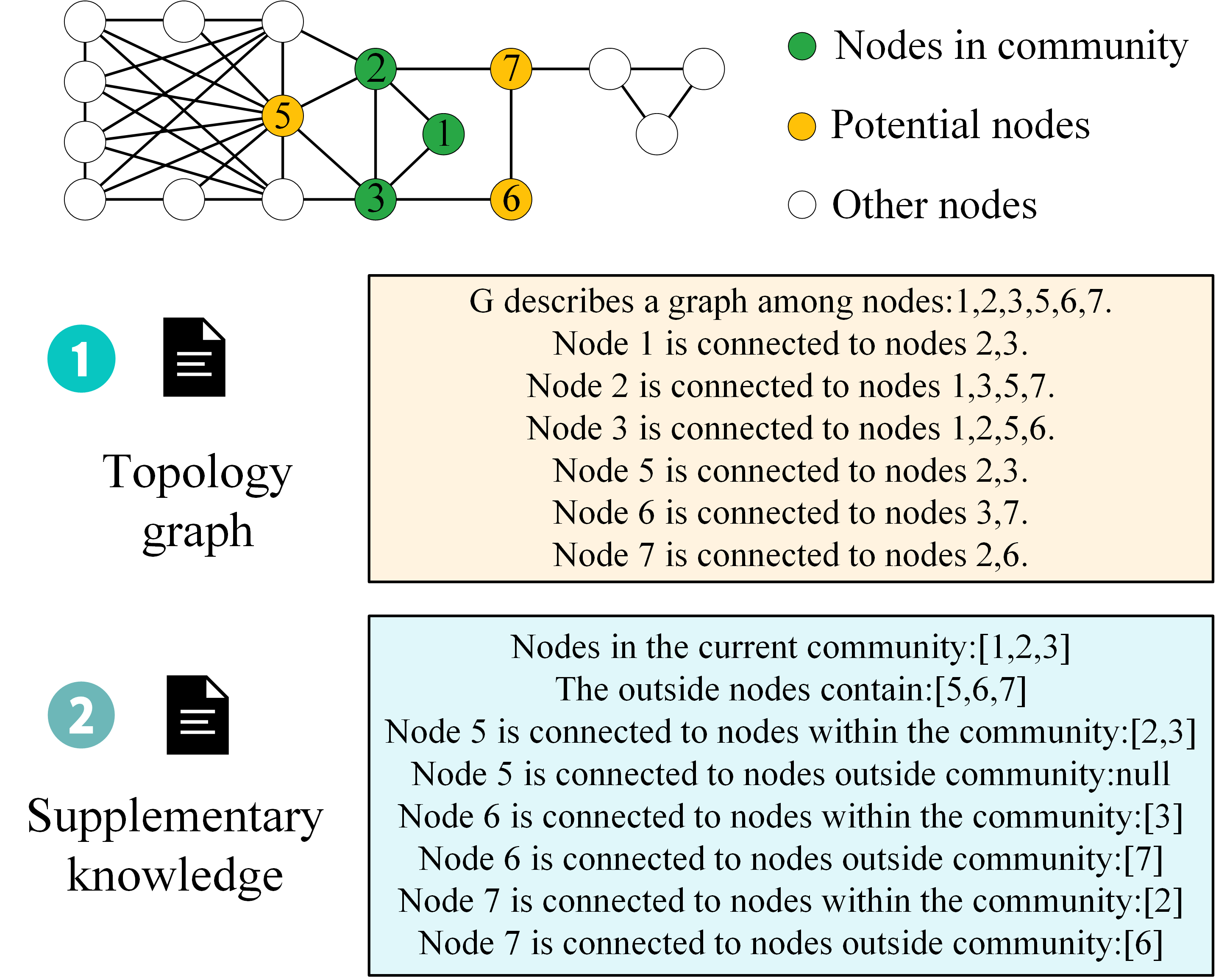} 
    \caption{ An toy example of graph encoding. The middle part illustrates the  text corresponding to topology graph.  The lower part illustrates the  text corresponding to  supplementary knowledge.}
    \label{fig:Graph encoding}
\end{figure}

The purpose of graph encoding is to convert graph structure data and information about the community $C$ into text that is easily understandable by GPT-3.5-turbo.
We design graph encoding method, called  ComIncident, for the community detection task. 
ComIncident includes two aspects: 
1) \textbf{Topology graph.}  The graph topology is represented by introducing the first-order neighbors of each node, as in Incident \cite{fatemi2023talk}.
2) \textbf{Supplementary knowledge}. We describe node information from a community perspective to enhance GPT-3.5-turbo's understanding of community structure.
It involves outlining which nodes are within the community, those that are outside, and the connections of potential nodes to both within and outside the community. 
The topological structure of local network  and the supplementary information are merged to form $gtxt$.
ComIncident is an improved version of Incident. Incident utilizes only the topology graph, whereas ComIncident incorporates both the topology graph and supplementary knowledge. 
To better understand ComIncident, an example is given in Figure \ref{fig:Graph encoding}.

For the node selection by GPT-3.5-turbo step (Section \ref{sec:Node selection by GPT-3.5-turbo}), the local network comprises nodes in $C \cup PN$ and the edges between them. 
For the node supplementation by GPT-3.5-turbo step (Section \ref{sec:Node supplementation by GPT-3.5-turbo}), the local network consists of  nodes in $C\cup N$ and edges between these nodes, where $N$ includes the first and second-order neighbors of community $C$.

\subsection{Node selection by GPT-3.5-turbo}
\label{sec:Node selection by GPT-3.5-turbo}

Node selection refers to choosing the most suitable node for community expansion, based on NSG (Node Selection Guide) prompt, graph text $gtxt$, and question Q$_{sel}$.
Graph text $gtxt$ is obtained by the graph encoding step (Section \ref{sec:Graph encoding}) with the local network, $C$, and $PN$, where the local network consists of the nodes in $C\cup PN$ and edges between these nodes.
Below, we introduce the NSG prompt  and question Q$_{sel}$, where the NSG prompt contains NSG\_1 and NSG\_2.

 \textbf{Prompt NSG\_1: \textit{The more an outside node is connected to other outside nodes, the higher the likelihood of its selection}}.
 
 The outside nodes provided to GPT-3.5-turbo are  potential nodes from the  potential nodes identification step (Section \ref{sec:Potential nodes identification}). This prompt could alleviate the seed-dependent problem and the free rider effect. The reason for the seed-dependent problem is low-degree node preference.
Take seed node 5 in Figure \ref{Seed_dependent_problem} 
and M method \cite{M} as an example to illustrate it. 
Nodes 1, 7, and 11 are  connected to nodes within the community [5]. 
The M method \cite{M} selects node 11, having the smallest degree, to join the community, ensuring the fewest external edges of the community.
With the incorporation of the prompt NSG\_1, nodes that connect to many other potential nodes are prioritized. Consequently, node 1 is selected over the mistakenly considered node 11.  
That is, prompt NSG\_1 helps to avoid bias towards low-degree nodes and thereby alleviates the seed-dependent problem.
The reason for the free rider effect is adding nodes that are sparsely connected to the community to the community. Nodes that are connected to multiple potential nodes indirectly maintain close connections to the community, which alleviates the free rider effect. Taking  Figure \ref{free_rider_effect} 
as an example, the community contains node 7, with nodes 4, 6, and 8 regarded as potential nodes. According to NSG\_1, nodes 6 and 8 have a higher probability of being selected. Node 4 is a node that may cause the free rider effect, and will not be selected.

\textbf{Prompt NSG\_2: \textit{Prioritize selecting outside nodes that are connected to multiple nodes within the community}}. 
The outside nodes mentioned in prompts are potential nodes from the potential nodes identification step (Section \ref{sec:Potential nodes identification}). 
This prompt could alleviate the community diffusion. 
The reason for the community diffusion is the insufficient consideration of the connections between nodes to be added to the community and the nodes within the community. Therefore, the prompt considers the connection between potential nodes and the community, thereby alleviating the community diffusion problem.
Taking Figure \ref{community_diffusion} 
as an example,  the community includes nodes 1 and 3, with nodes 2, 5, 7, 13, 15, and 18 regarded as potential nodes, where nodes 13 or 18 may lead to community diffusion. 
According to NSG\_2, nodes 2 and 5 are more likely selected, rather than mistakenly selecting nodes 13 or 18.

The NSG\_1 considers the closeness of the node and potential nodes that are highly likely to be added into the community as ComGPT proceeds. The NSG\_2 focuses on the closeness of the node and nodes within the community. These two prompts align with the characteristics of communities where nodes in the community are tightly connected, thus helping GPT-3.5-turbo understand the characteristics of the nodes to be selected. 
To guide GPT-3.5-turbo to  provide relevant answers, the question  Q$_{sel}$ is designed: 

``\textit{You're doing local community detection. Based on the graph data and prompt, please select a node that you think is most likely to belong to the current community $C$ for community expansion. Provide a detailed explanation.}".

Based on the above prompts, the graph text $gtxt$ (Section \ref{sec:Graph encoding}), and the question Q$_{sel}$, GPT-3.5-turbo selects the node $v$ from potential nodes.

\subsection{Node supplementation by GPT-3.5-turbo} 
\label{sec:Node supplementation by GPT-3.5-turbo}

Node supplementation involves evaluating whether to continue the expansion process or terminate it. 
If continuing, it also determines which additional nodes should be selected. 
Next, we introduce  NSU (Node SUpplementation) prompt and question Q$_{sup}$ in turn.

NSU prompt is based on the definition of the community and whether these nodes can maintain the cohesion of the community, as follows:

``\textit{Definition of community: nodes in the same community are tightly connected, while the nodes in different communities are sparsely connected. The more a node connects with nodes within a community, the more it can increase the community's cohesion. Conversely, the more it connects with nodes outside the community, the more it can decrease the community's cohesion. If you choose to add a node to the community, it should make the connections within the community tighter.}".

The question Q$_{sel}$ for supplementing nodes is designed as:

``\textit{Please analyze whether these nodes should be added to the community $C$. The probability of not adding nodes is higher. But it doesn't mean you always refuse to add nodes. If you think there is a suitable node, please output its node number.}".

Based on the NSU prompt, the text obtained by the graph encoding step (Section \ref{sec:Graph encoding}), and the question Q$_{sel}$, GPT-3.5-turbo determines whether any of the nodes warrant inclusion in the community. 
If so, it returns the node's number; otherwise, it returns `null'.

ComGPT aims to explore the application of LLMs to local community detection. Specifically, it skillfully combines LLMs with the seed expansion algorithm. Different from traditional seed expansion methods that rely on predefined and rigid heuristic rules, ComGPT leverages the implicit semantic reasoning capability of LLMs to guide the seed expansion process.  
Notably,  we do not simply replace a certain rule of the seed expansion algorithm with LLMs. 
Instead, ComGPT provides the LLM with guiding principles rather than explicit, step-by-step rules.
This leverages the LLM's  inherent reasoning capability  to grasp the nuanced semantics of concepts like `tightness' and `community membership'. Consequently, the LLMs can  be prompted to reason about diverse network structures.
\subsection{Complexity analysis}

The  time complexity analysis is as follows. Assuming there are $n$  nodes in the local community and the average degree of nodes is $d$. 
The potential node identification step involves traversing the neighbors of nodes within the local community and calculating modularity M, with a time complexity of O($nd^2$). 
The graph encoding step requires generating textual descriptions for the current community and potential nodes, with a time complexity proportional to the number of nodes and their degrees, i.e., O($nd$). The core of the node selection and node supplementation steps involves calling the GPT-3.5-turbo API. Since there is no literature defining the time complexity of GPT-3.5-turbo API calls, this portion of the time complexity remains unestimated. ComGPT repeats the above steps $n$ times to get the final community. 
Excluding the GPT-3.5-turbo API calls, the time complexity of ComGPT is O($n^2d^2$).

The space complexity analysis is as follows.
The potential node identification step stores local community, neighbors of local community, and   $k$ potential nodes. In ComGPT, $k$ is treated as a constant with a default value of 10; therefore, the space complexity of this step is O($n+nd+k$). 
The graph encoding step requires storing the current community along with its potential nodes and the connections between nodes, resulting in a space complexity of O($nd+kd$). 
The node selection and node supplementation steps only store the response results from GPT-3.5-turbo, with a space complexity of O($1$). Therefore, the overall space complexity of ComGPT is O($nd+kd$).

\section{Experiments}
\label{sec:Experiments}

In this section, we first introduce the experimental settings,
including datasets, baselines, implementation details, and evaluation metrics. Then, we give the experimental results and corresponding analyses.

\subsection{Experimental settings}

\subsubsection{Datasets}


\begin{table}
    \centering
    \caption{Statistics of the datasets.}
    \begin{tabular}{c c c c c c c}
        \toprule
        Networks & Nodes & Edges & $\bar{d}$ & $|C|$ & $\mu$ & $\bar{Q}$\\
        \midrule
        Dolphins & 62 & 159 & 5.13 & 31 & 0.0383 & 0.3734\\
        Football &  115 & 613 & 10.66 & 9.58 & 0.3638 & 0.5539\\
        Polbooks &  105 & 441 & 8.4 & 35 & 0.1812 & 0.4149\\
        Amazon & 334863 & 925872 & 5.53 & 30.23 & 0.1216 & 0.4441\\
        DBLP &  317080 & 1049866 & 6.62 & 53.41 & 0.2147 & 0.4453\\
        \bottomrule
    \end{tabular}
    \label{datasets}
\end{table}

Five real-world  datasets, including Dolphins \cite{lusseau2003emergent}, Football \cite{girvan2002community}, Polbooks \cite{krebs2004political}, Amazon \cite{yang2012defining}, and DBLP \cite{yang2012defining} are utilized to test the performance of algorithms.  Table \ref{datasets} shows the detailed information of five real networks, including the number of nodes, the number of edges, the average degree $\bar{d}$, the average community size $|C|$, the mixing parameter $\mu$ \cite{DMF}, and the modularity $Q$ \cite{Q}.

The Dolphins network is based on observations of associations between pairs of dolphins, where edges represent associations between dolphins.
The Football network is based on college football matches. Nodes represent college football teams, communities correspond to different leagues, and edges correspond to games played between them. 
The Polbooks network is a co-purchasing network for books. Each node in the dataset represents a book about US politics. An edge between two books indicates that they are often purchased together by customers. 
The Amazon network represents product co-purchasing relationships. Nodes representing products and edges indicate which products are frequently co-purchased.
The DBLP network is a co-authorship network. Here, nodes represent authors and edges indicate that two authors have co-authored a paper. 







\subsubsection{Baselines}
To evaluate the performance of ComGPT, we compare it with M method \cite{M}, R method \cite{clauset2005finding}, Louvain \cite{Louvain}, DMF\_M \cite{DMF}, DMF\_R \cite{DMF}, Leiden \cite{Leiden}, LCDNN \cite{LCDNN}, LS \cite{LS}, and FLMIG \cite{FLMIG}, introduced as follows:
\begin{enumerate}
\item  M method \cite{M} and R method \cite{clauset2005finding}. The M method and R method are greedy algorithms based on modularity. The given node constitutes the initial local community at first.  M method (R method) iteratively selects and adds the node that maximizes the modularity  $M$($R$)  to the local community until no node will decrease modularity $M$($R$).

\item Louvain \cite{Louvain}. The Louvain first treats each node as an individual community and iteratively adjusts node assignments to maximize modularity until stability is reached. Then, it aggregates communities into new nodes, constructs a simplified graph, and repeats the optimization process until modularity converges.

\item DMF\_M and DMF\_R\cite{DMF}. DMF\_M and DMF\_R analyze the formation process of the local community and divide it into three stages. Each stage corresponds
to a different dynamic membership function.

\item Leiden \cite{Leiden}. Leiden is a derivative version of Louvain. Unlike Louvain, Leiden ensures well-connected communities through a refinement step. It then contracts communities into new nodes, constructs a simplified graph, and repeats the process until convergence, achieving more stable partitions.

\item LCDNN \cite{LCDNN}. LCDNN relies on NGC nodes. The nodes are added to the local community based on the fuzzy relations between them and their NGC nodes. At each time, if the largest fuzzy relation is no less than half of the average fuzzy relation of the current local network, one node with the largest fuzzy relation is added to the local  community.

\item LS \cite{LS}. LS is built on the notion of local dominance, where low-degree nodes are assigned to the basin of influence of high-degree nodes. Utilizing this principle, LS identifies community centers based on node degree and other local information. Finally, it divides communities according to the identified community centers.

\item FLMIG \cite{FLMIG}. FLMIG is also a derivative version of Louvain. FLMIG enhances the Louvain Prune heuristic using an iterated greedy (IG) framework to maximize modularity in non-overlapping communities.

\end{enumerate}
For baselines that require parameter settings, we select the optimal parameter within the given range as the final value. For Louvain, the parameter that controls the size of the communities takes values within  [0.3, 0.5, 0.8, 1, 1.5]. For Leiden,  the maximum total size of nodes in a community is tuned within  [0, 50, 100, 150, 200]. For FLMIG, the parameters are set as explicitly provided by the authors.

\subsubsection{Implementation Details} ComGPT uses gpt-3.5-turbo-0125. The number of potential nodes $k$ is set to 10.
 All experiments are  conducted under Windows 10 operating system (CPU: 11th Gen Intel(R) Core(TM) i7-11700K @ 3.60GHZ, Memory: 32 GB.). Due to the monetary costs of GPT-3.5-turbo-0125, we execute ComGPT twice for each starting seed node to handle GPT hallucinations. The code is released on the author’s website\footnote{https://github.com/cassius0301/ComGPT}.

\subsubsection{Evaluation metrics}

The evaluation metrics Jaccard \cite{Jaccard}  and  Fscore \cite{F1} are used to evaluate the detected communities. Their formulas are as follows.

\begin{equation} \label{eql:J}
\text{Jaccard} = \frac{\left| C_{\text{Found}} \cap C_{\text{True}} \right|}{\left| C_{\text{Found}} \cup C_{\text{True}} \right|},
\end{equation}

\begin{equation} \label{eql:F}
\text{Fscore} = \frac{2 \cdot \text{Precision} \cdot \text{Recall}}{\text{Precision} + \text{Recall}},
\end{equation}
\begin{equation} \label{eql:R}
\text{Recall} = \frac{\left| C_{\text{Found}} \cap C_{\text{True}} \right|}{\left| C_{\text{True}} \right|},
\end{equation}

\begin{equation} \label{eql:P}
\text{Precision} = \frac{\left| C_{\text{Found}} \cap C_{\text{True}} \right|}{\left| C_{\text{Found}} \right|},
\end{equation}
where $C_{\text{Found}}$ is the local community that the algorithm discovers, and $C_{\text{True}}$ is the real local community to which the given starting node belongs. 
The values of Fscore and Jaccard are between 0 and 1. The larger value implies a better algorithm performance. For an overlapped seed node, all the communities it belongs to are integrated as a real local community.

\subsection{Experimental results}

\begin{table*}
\centering
\caption{Experimental results on five datasets. On each dataset, the highest Fscore (Jaccard) is highlighted in bold, the second highest one is highlighted by underline, and the third highest one is highlighted with a superscript asterisk. }
\begin{tabular}{cccccccccccc}
\toprule
Dataset & Metrics & M & R  & Louvain & DMF\_M & DMF\_R & Leiden & LCDNN & LS & FLMIG & ComGPT \\
\midrule
\multirow{2}{*}{Dolphins} & Jaccard & 0.3370 & 0.3128 & 0.4705 & 0.6270 & 0.6320 & 0.5307 & 0.6438*  & \textbf{0.7735} & 0.5359 & \underline{0.7035} \\
& Fscore  & 0.4673 & 0.4478 & 0.5872 & 0.7254 & 0.7241 & 0.6320 & 0.7558*  & \textbf{0.8385} & 0.6517 & \underline{0.7684} \\
\midrule
\multirow{2}{*}{Polbooks} & Jaccard & 0.4266 & 0.3986 & 0.6403 & 0.6449 & 0.6437 & 0.6374 & \underline{0.6673}  & \textbf{0.6919} & 0.6336 & 0.6468* \\
& Fscore  & 0.5184 & 0.4960 & 0.7229 & 0.7294 & 0.7285 & 0.7237  & \underline{0.7463}  & \textbf{0.7764} & 0.7191 & 0.7320* \\
\midrule
\multirow{2}{*}{Football} & Jaccard & 0.6117 & 0.6134 & 0.8516* & \underline{0.8546} & \textbf{0.8579} & 0.7762  & 0.7547  & 0.2406 & 0.5596& 0.7571 \\
& Fscore  & 0.6706 & 0.6699  & \underline{0.8890} & 0.8881* & \textbf{0.8893} & 0.8361  & 0.8255  & 0.3593 & 0.6798 & 0.8103 \\
\midrule
\multirow{2}{*}{Amazon}   & Jaccard & 0.6961 & 0.6771 & 0.1917 & 0.7633 & \underline{0.7686} & 0.4741  & 0.7665*  & 0.6506 & 0.0479 & \textbf{0.7734} \\
& Fscore  & 0.7706 & 0.7554 & 0.2499 & 0.8353 &  0.8361* & 0.5940 & \underline{0.8372}  & 0.7420 & 0.0642 & \textbf{0.8421} \\
\midrule
\multirow{2}{*}{DBLP}     & Jaccard & \underline{0.3239} & 0.3207* & 0.1786 & 0.2945 & 0.2928 & 0.2076  & 0.2208  & 0.0650 & 0.0074 & \textbf{0.3259} \\
& Fscore  & \underline{0.4185} & \underline{0.4185} & 0.2657 &  0.4032 & 0.4009 & 0.3187  & 0.3152  & 0.0850 & 0.0144 & \textbf{0.4359} \\
\bottomrule
\end{tabular}
\label{tab:algorithm experimental results.}
\end{table*}

\subsubsection{Results on networks}
 For the Dolphins, Polbooks, and Football datasets, each node is taken as the starting node to obtain the local community.
 For the Amazon and DBLP datasets, we randomly select 200 nodes from the top 5,000 communities as the starting nodes.
 The average values of Fscore and Jaccard for the starting nodes are  shown in Table \ref{tab:algorithm experimental results.}. 


Table \ref{tab:algorithm experimental results.} shows that ComGPT exhibits the best performance on the Amazon and DBLP datasets, ranking second and third places on the Dolphins and Polbooks datasets, respectively.  On the Football dataset, ComGPT does not perform well.
The reasons for the inconsistent performance of ComGPT across datasets are explained as follows.
ComGPT achieves the highest performance on Amazon and DBLP. As shown in Table \ref{datasets}, these datasets have low mixing parameters (0.1216 and 0.2147) and high modularity (0.44), indicating distinct community structures. This structure enables GPT-3.5-turbo to easily distinguish nodes inside and outside communities, thus achieving optimal results.
For Polbooks and Dolphins, the modularity is lower than that of Amazon and DBLP, indicating weaker community structures. This increases the difficulty for GPT-3.5-turbo to accurately identify community members, resulting in some misidentified nodes and preventing ComGPT from achieving optimal performance. For Football, it has a high mixing parameter (0.3638), meaning numerous inter-community connections. These connections may cause GPT-3.5-turbo to merge multiple real communities into one, leading to ComGPT's poor performance.


LS performs well on datasets like Dolphins and Polbooks because it can accurately identify local leaders. This reliance on local leaders becomes a weakness in networks with strong inter-community connections, such as the football dataset. In these cases, a node identified as a local leader might actually be a follower of a more global leader, causing the algorithm to perform poorly.
Louvain and its improved versions, such as Leiden and FLMIG,  underperform on Amazon and DBLP datasets, as it tends to merge multiple distinct real communities into a single large community.
Except for the DBLP dataset, DMF\_M, DMF\_R, and LCDNN perform better than the M method and R method. The reason is that the latter two methods maximize local modularity ($M$ and $R$), making them sensitive to the choice of seed nodes, known as seed-dependent problem. In contrast, DMF\_M, DMF\_R, and LCDNN implement various strategies to alleviate this problem. 
Specifically, DMF\_M and DMF\_R utilize membership functions to select nodes with large degrees, while LCDNN expands the community based on the fuzzy relations between nodes and their NGC nodes.

ComGPT aims to explore the application of LLMs to local community detection. The baselines are mature methods, optimized and tuned for excellent performance. In contrast, ComGPT, an emerging exploratory method based on LLMs, is still in its early stages and requires further optimization to reach its best performance. Its value lies in revealing the potential of LLMs in local community detection.

\subsubsection{Ablation study}

\begin{figure}[!t]
\centering
\subfloat[\small Dolphins]
{
\includegraphics[width=0.44\columnwidth]{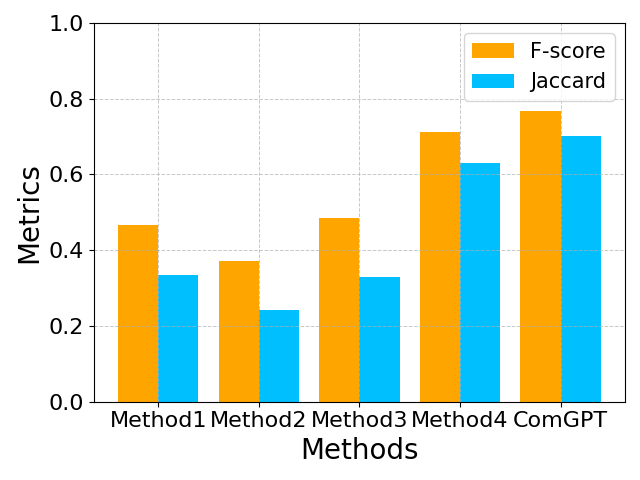}}
\hspace{0mm}
\subfloat[\small Polbooks]
{
\includegraphics[width=0.44\columnwidth]{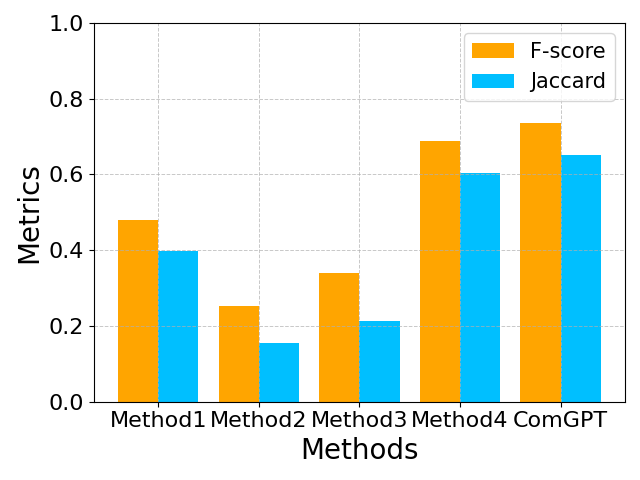}
}
\caption{Ablation study.}
\label{fig:Ablation study}
\end{figure}

To explore the necessity of GPT-3.5-turbo, SK (Supplementary knowledge in graph coding), NSG, and NSU prompts in ComGPT, we conduct ablation experiments on the Dolphins and Polbooks datasets with 50 randomly selected seed nodes.
Four simplified versions of ComGPT are constructed as follows: 1) Method 1 without GPT-3.5-turbo. The first node in $PN$ is selected for community expansion until $PN$ is empty; 2) Method 2  without SK; 3) Method 3  without NSG prompt; and 4) Method 4  without NSU prompt.

The simplified version of ComGPT may add duplicate nodes to the detected communities. When calculating Fscore and Jaccard, it is necessary to obtain the intersection between the detected community and the true community. Here, duplicate nodes in the detected community are regarded as a single node. Additionally, retain the duplicate nodes in the detected community. To prevent cyclic selection of duplicate nodes, community expansion is terminated once the number of duplicate nodes exceeds one-third of the number of nodes in the community.

Figure \ref{fig:Ablation study} shows the results of Methods 1, 2, 3, 4, and ComGPT.
Overall, ComGPT outperforms Methods 1, 2, 3, and 4, indicating that GPT-3.5-turbo, SK, NSG, and NSU prompts play a crucial role in ComGPT. ComGPT is better than Method 1 in terms of Fscore and Jaccard, which demonstrates the importance of GPT-3.5-turbo. The reason ComGPT outperforms Method 2 is that its graph coding incorporates community knowledge. Compared with graph encoding methods that contain only the graph topology, supplementing community knowledge enables the GPT-3.5-turbo to better understand the community structure and thus make more accurate judgments. ComGPT also outperforms Method 3 and 4, suggesting that NSG and NSU prompts can enhance its performance in local community detection.

\subsubsection{Prompt comparisons}

\begin{table}
  \centering
  \caption{Description of the prompts.}
  \begin{tabular}{lp{6cm}}
    \toprule
    Prompts & Description    \\
    \midrule
    Few-shot & It is given a few demonstrations of the task 
    \newline at inference time as conditioning.  \\
    Zero-shot & No demonstrations are allowed and only an instruction in natural language is given to the model.  \\
    CoT & Let's think step by step.  \\
    BaG & Let’s construct a graph with the nodes and edges first.  \\
    \bottomrule
  \end{tabular}
  \label{tab:prompt introduction}
\end{table}

\begin{figure}[!t]
\centering
\subfloat[\small Dolphins]
{
\includegraphics[width=0.44\columnwidth]{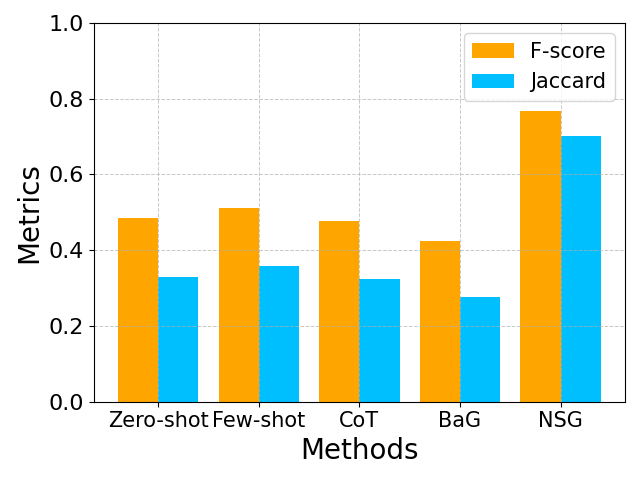}}
\hspace{0mm}
\subfloat[\small Polbooks]
{
\includegraphics[width=0.44\columnwidth]{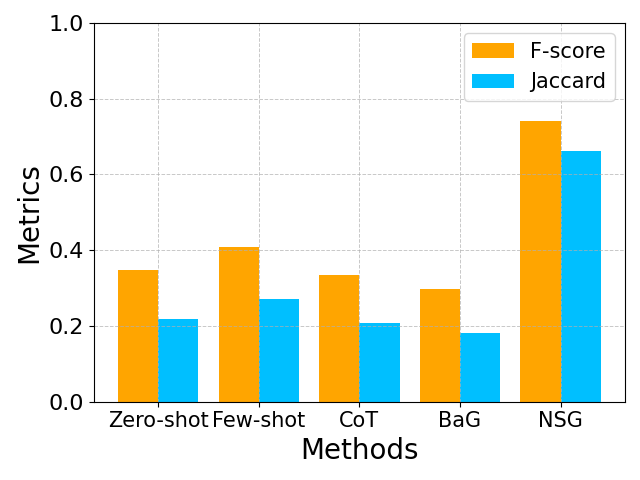}
}
\caption{Results of different prompts.}
\label{fig:Promptcomparison}
\end{figure}

To test the performance of the NSG prompt within ComGPT, we replaced the NSG prompt with four typical prompt approaches: Zero-shot \cite{brown2020language}, Few-shot \cite{brown2020language}, CoT \cite{wei2022chain}, and BaG \cite{wang2024can} prompts, as shown in Table \ref{tab:prompt introduction}. We conducted comparative experiments by randomly selecting 50 seed nodes in the Polbooks and Dolphins datasets.
Detailed experimental results are shown in Figure \ref{fig:Promptcomparison}. From these results, we derive the following conclusions:

Example samples are somewhat helpful to GPT-3.5-turbo. Based on the experimental results of Zero-shot and Few-shot, adding sample examples led to performance improvements in terms of Fscore and Jaccard on the Dolphins and Polbooks datasets. This indicates that providing examples for the GPT-3.5-turbo is indeed effective. Appropriate examples can improve performance.

Advanced prompts may have limited performance improvement or even counterproductive effects on complex graph reasoning tasks.  For instance, advanced prompts like CoT can sometimes introduce more points of failure, as each step must be correct to ensure a successful final decision. It will increase the risk of error accumulation, especially in complex graph-based tasks. Consequently, advanced prompts may not always improve performance.

Compared to other prompts, NSG brings the highest improvement. Our designed prompt NSG achieve the best performance in terms of Fscore and Jaccard on Dolphins and Polbooks datasets. This is because the design of CoT and BaG is oriented towards universal graph reasoning problems so that no relevant guidance or knowledge is utilized in dealing with local community detection. NSG is oriented to the field of local community detection and utilizes the guidance in local community detection. 
﻿


\subsubsection{Effect of NSG\_1 and NSG\_2 on ComGPT}
\begin{figure}[!t] 
    \centering
    \includegraphics[width=0.8\linewidth]{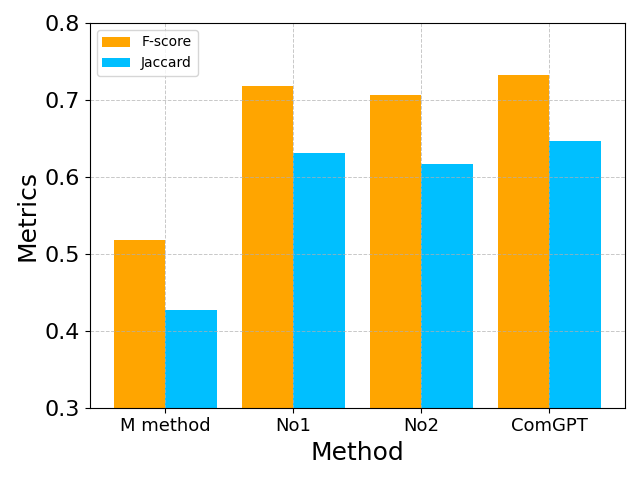} 
    \caption{Effect of NSG\_1 and NSG\_2 on ComGPT.}
    \label{fig:nsg}
\end{figure}

We explore the effect of NSG\_1 and NSG\_2  on ComGPT  on the polbooks dataset.
 In ComGPT, the NSG\_1 prompt handles the seed-dependent problem and free rider effect, while the NSG\_2 prompt tackles community diffusion (Section \ref{sec:Node selection by GPT-3.5-turbo}). 
 To verify whether ComGPT alleviates these shortcomings, we remove NSG\_1 and NSG\_2 from ComGPT to obtain methods named ``No1" and ``No2", respectively. 
 Result of No1, No2, ComGPT, and M Method on Polbooks is shown in Figure \ref{fig:nsg}. 
 Figure \ref{fig:nsg} shows that ComGPT, which combines the M method and GPT-3.5-turbo, outperforms the M method alone, indicating that ComGPT mitigates these shortcomings faced by the M method. ComGPT's superiority over No1 indicates that the first prompt mitigates seed-dependent problem and free-rider effect, while its superiority over No2 shows that the second prompt mitigates community diffusion.
\subsubsection{Effect of the number of potential nodes on ComGPT}

\begin{figure}[!t] 
    \centering
    \includegraphics[width=0.8\linewidth]{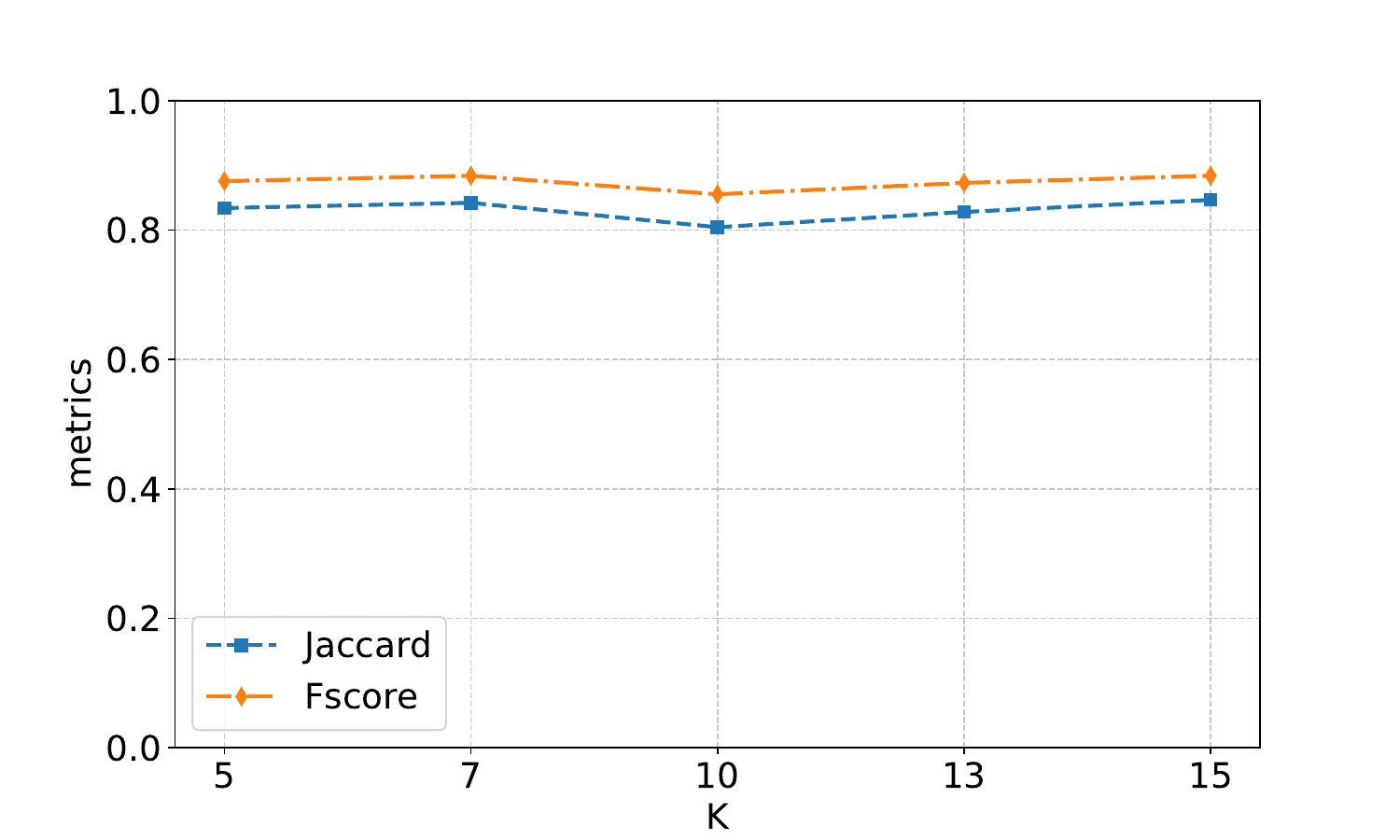} 
    \caption{Effect of $k$ on ComGPT.}
    \label{fig:K}
\end{figure}

We explore the effect of the number of potential nodes, denoted by $k$, on ComGPT  on the Football dataset, with 50 seed nodes selected randomly. The values of $k$ are 5, 7, 10, 13, and 15.
Figure \ref{fig:K} shows the result of ComGPT under different $k$. 
As $k$ increases, the performance of ComGPT remains generally stable, though with some fluctuations.  
Fluctuations are relatively low across different values, except for $k$ = 10, which exhibits higher variability due to the randomness of GPT-3.5-turbo output.
To confirm this, we repeat experiments for $k$ = 10 and the results are nearly identical to those obtained with other values of $k$.
The above analysis suggests that the performance of ComGPT is less affected by the parameter $k$.

\subsubsection{Analysis of GPT-3.5-turbo API Runtime in ComGPT }
\begin{table}[!t]
    \centering
    \caption{Execution Times on node 1 of the Football dataset.}
    \label{tab:execution_time}
    \begin{tabular}{ccccc}
        \toprule
        & \multicolumn{2}{c}{\textbf{GPT-3.5-turbo Response Time}} & \multicolumn{2}{c}{\textbf{Other Execution Time}} \\
        \cmidrule(lr){2-5}  
        \textbf{} & \textbf{t(s)} & \textbf{Ratio(\%)} & \textbf{t(s)} & \textbf{Ratio(\%)} \\
        \midrule
        1 & 133.4220 & 99.9441 & 0.0745 & 0.0559 \\
        2 & 285.5551 & 99.9755 & 0.0699 & 0.0245 \\
        3 & 150.5789 & 99.9595 & 0.0609 & 0.0405 \\
        4 & 298.0469 & 99.9805 & 0.0579 & 0.0195 \\
        5 & 128.2163 & 99.9540 & 0.0589 & 0.0460 \\
        \textbf{Average} & 199.1678 & 99.9677 & 0.0644 & 0.0323 \\
        \bottomrule
    \end{tabular}
\end{table}

To analyze the runtime  of calling the GPT-3.5-turbo API in ComGPT, we selected node 1 of the Football network as the starting node. 
We record the execution time of ComGPT at this starting node over five runs, including the GPT-3.5-turbo response time, the ratio of GPT-3.5-turbo response time in ComGPT, the local computation time, the proportion of local computation time in ComGPT, and the average values of these results. The experimental results are shown in Table \ref{tab:execution_time}. 

The experimental results show that the response time of GPT-3.5-turbo accounts for more than 99.9\% of the ComGPT's execution time, indicating that the overall execution efficiency is mainly limited by GPT-3.5-turbo's response speed rather than other computational processes.
ComGPT needs to call GPT-3.5-turbo and wait for its response. This process is influenced by various factors, such as network conditions and server load, making it difficult to predict the exact response time. 

\subsubsection{Scalability}

To evaluate the scalability of ComGPT, we randomly sample subgraphs with different sizes from the Amazon dataset and test ComGPT on these extracted subgraphs. 
We extract subgraphs containing 20\%, 40\%, 60\%, 80\%, and 100\%  of the nodes from the Amazon dataset, respectively.
We randomly select 100 nodes as the starting nodes from each subgraph. The average values of Precision, Recall, Fscore, Jaccard, runtime, and number of tokens are shown in Table \ref{tab:scalability}.

Table \ref{tab:scalability} shows that, as the percentage of nodes increases, the Recall, F-score, and Jaccard of ComGPT improve. This is because, as the number of nodes and edges in the network increases, the community structure becomes more obvious, and the communities detected by ComGPT gradually become more complete.
Furthermore, as the percentage of nodes increases, we observe that the Precision fluctuates while the Recall consistently increases. This growth in Recall indicates that the size of the detected communities also increases with the network size, covering more ground-truth members. The fluctuation in Precision suggests that this expansion is not always perfectly accurate. 
The total runtime and number of tokens of ComGPT increase as the network grows. As the percentage of nodes increases, the average degree of the nodes in the network rises. Since the time complexity is proportional to the average degree, the runtime increases as the percentage of nodes increases. As the percentage of nodes increases, the size of the detected communities also increases, leading to a higher number of required API calls.

\begin{table}[htbp]
\centering
\caption{Scalability of ComGPT on the Amazon dataset.}
\label{tab:scalability}
\begin{tabular}{cccccc}
\toprule
 & 20\% & 40\% & 60\% & 80\% & 100\% \\
\midrule
Precision & 0.9930 & 0.9657 & 0.9609 & 0.9520 & 0.9332 \\
Recall & 0.1497 & 0.3121 & 0.4779 & 0.6973 & 0.8601 \\
Fscore & 0.2452 & 0.4322 & 0.5954 & 0.7729 & 0.8646 \\
Jaccard & 0.1490 & 0.3038 & 0.4594 & 0.6719 & 0.8033 \\
Time (s) & 28.63 & 80.98 & 84.95 & 181.97 & 352.07 \\
Tokens  & 383.14 & 1344.21 & 4026.44 & 11431.69 & 12386.43 \\
\bottomrule
\end{tabular}
\end{table}

\section{Related work}
\label{sec:Related work}
In this section, we briefly review the related work on community detection and the application of Large Language Models (LLMs) in graph analysis.

\subsection{Community detection}
\label{chap:Community detection}
Community detection aims to identify groups of closely interconnected nodes.
According to whether they utilize global information of the network, existing community detection methods are classified into two types: global community detection and local community detection. 

\subsubsection{Global community detection}
 Global community detection depends on the network's global information to discover all communities. In recent years, researchers have investigated various algorithms \cite{9511798}, such as the Label Propagation Algorithm (LPA) \cite{FBLD}, Non-negative Matrix Factorization (NMF) \cite{wang2017community}, random walk based   techniques \cite{randomwalk}, and so on.
 Raghavan et al. introduced the Label Propagation Algorithm (LPA) \cite{raghavan2007near}. In the LPA, each node iteratively updates its label to the label with the highest frequency. Communities are divided based on the distribution of labels.
Wang et al. developed Non-negative Matrix Factorization (NMF)-based algorithms tailored for undirected networks, directed networks, and compound networks \cite{wang2011community}. The algorithm determines the membership of nodes to communities by decomposing the adjacency matrix.
Meo et al. proposed that edge weights enhance community detection and developed the WERW-Kpath \cite{DEMEO2013648}. It consists of two steps: (1) selecting a source node based on the degree of the node and the number of nodes in the network, and (2) choosing an unvisited edge from current node with probability proportional to its weight.

The above studies require complete network data to identify all communities. In contrast, our work focuses on determining the community of a specific node using only local information, thereby fundamentally distinguishing it from global studies.

\subsubsection{Local community detection}
Local community detection focuses on discovering the community to which the seed node belongs \cite{DBLP:journals/tkdd/NiYLZ24,9913716}, without involving the entire network. Compared to global community detection, it is more efficient and avoids unnecessary computational overhead. 
A variety of methods,  like $k$-core decomposition \cite{DBLP:conf/sigmod/CuiXWW14} and seed expansion algorithms \cite{clauset2005finding}, are developed for detecting local communities.
Seed expansion algorithms expand communities by selecting nodes with high scoring function values \cite{moradi2014local}.
For example, the R method \cite{clauset2005finding} and M method \cite{M} choose nodes based on the  local modularity R and M, respectively.
The seed expansion algorithms often face the seed-dependent problem \cite{10.1145/2623330.2623621}, community diffusion, and free rider effect \cite{10.1145/1835804.1835923,10.1145/1321440.1321526,10.1145/2487575.248764}. 
To address the seed-dependent problem, researchers developed different strategies such as identifying core node sets \cite{LCDMD,CAELCD}, finding the nearest nodes with greater centrality \cite{LCDNN},  alternating strategy of strong fusion and weak fusion \cite{ASFWF}, and designing membership functions with a bias towards nodes with large degrees \cite{DMF}.
Differing from the above studies, our work alleviates the aforementioned shortcomings by utilizing LLMs.



\subsection{LLMs in graph analysis}

\subsubsection{Application of LLMs in graph reasoning}

In recent years, researchers have begun to explore the potential of LLMs for applications in the areas of graph reasoning and graph machine learning. Wang et al. designed the NLGraph dataset for evaluating the graph reasoning capabilities of LLMs and demonstrated that LLMs do have graph reasoning abilities \cite{wang2024can}. 
Zhang et al. designed the Graph-ToolFormer framework to handle graph reasoning tasks \cite{zhang2023graphtoolformerempowerllmsgraph}. It enabled LLMs to self-guide using ChatGPT-enhanced prompts.
Chen et al. utilized LLMs for the node classification task and proposed two approaches: LLMs-as-Enhancers and LLMs-as-Predictors \cite{10.1145/3655103.3655110}. The former leverages LLMs to enhance the node’s text attributes with their massive knowledge and then generates predictions through GNNs. The latter attempts to directly employ LLMs as standalone predictors. Chen et al. proposed a new label-free pipeline LLM-GNN to leverage LLMs for annotation, providing training signals on GNN for further prediction \cite{chen2024labelfreenodeclassificationgraphs}. Specifically, LLMs are leveraged to annotate a small portion of nodes and then GNNs are trained on LLMs’ annotations to make predictions for the remaining large portion of nodes.

\subsubsection{Graph encoding in LLMs}

In the field of LLMs in graph machine learning research, scholars have focused on how to improve the ability of LLMs in processing graph data.
Guo et al. emphasized the design and improvement of graph encoding techniques, which critically affect LLMs' interpretation of input data and its subsequent output \cite{guo2023gpt4graph}.
Therefore, Adjacency and Incident \cite{fatemi2023talk} are designed to encode graph structured data into text for LLMs. Adjacency uses integer node encoding and parenthesis edge encoding. Incident uses integer node encoding and incident edge encoding. 
Existing graph coding methods primarily represent graph topology, which is often insufficient for specific tasks, such as community detection. 
Unlike existing approaches, our work introduces a novel graph encoding method tailored for the community detection task.

\subsubsection{Prompts in LLMs}

PROMPT engineering is an approach to increase model inputs through prompts used for specific tasks so that LLMs are better adapted to new tasks \cite{sahoo2024systematic}. The prompt is widely used in computer vision \cite{bahng2022exploringvisualpromptsadapting}, recommender systems \cite{10.1145/3655103.3655110}, and other fields. This approach does not require modification of model parameters, but rather significantly enhances the model's ability to process downstream tasks based on task-specific prompts. 
Various prompts are designed for specific tasks to make LLMs better adapt to these tasks \cite{sahoo2024systematic}, such as the general CoT (Chain-of-Thought) prompt \cite{wei2022chain} and the BaG (Build a Graph) prompt \cite{wang2024can}. CoT aims to enable LLMs to solve complex problems through step-by-step reasoning. Its core idea is to mimic human thought processes by reasoning incrementally rather than directly generating the final answer. BaG encourages LLMs to map the textual descriptions of graphs and structures to grounded conceptual spaces before tackling the specific problem through a one-sentence instruction.
However, these prompts do not incorporate the domain knowledge required for local community detection and are unsuitable for detecting community structures within networks. 
This paper designs prompts suitable for local community detection tasks, which are different from the above prompts.

\section{Conclusion}
\label{sec:Conclusion}
We explored using GPT-3.5-turbo to assist in seed expansion algorithm and developed a GPT-guided methodology, termed ComGPT. 
Additionally,  a graph encoding method suitable for local community detection is explored,  which enhances the existing graph encoding by incorporating community knowledge.
Furthermore, the NSG prompt is designed to enhance LLMs' understanding of community domain knowledge.
Experimental results confirm that  the designed graph encoding method and prompts help GPT-3.5-turbo understand the domain knowledge and detect local community structure.

Although ComGPT utilizes a local expansion strategy to minimize token consumption, the size of local communities in large-scale networks may  be so large that the input information exceeds the token limits of LLMs. 
In the future, we will explore scalable local community detection and information compression methods, aiming to ensure that when the community size is large, the information input to LLMs remains within their token limits. Besides, to explore the potential of the proposed idea, we  plan to extend it to other tasks for 2D/3D multi-media data \cite{CAD,EAT}.
\ifCLASSOPTIONcompsoc
  \section*{Acknowledgments}
\else
  \section*{Acknowledgment}
\fi

This work was supported by the National Natural Science Foundation of China [No.62572002, No.62272001 and No.62206004], Natural Science Foundation of Anhui Province of China [No.2508085MF159], Hefei Key Technology R\&D “Champion-Based Selection” Project [No.2023SGJ011], and Guangdong Provincial Key Laboratory of Novel Security Intelligence Technologies [No.2022B1212010005].

\bibliography{ComGPT_ref}
\bibliographystyle{IEEEtran}

\vspace{-1.15 cm}
\begin{IEEEbiography}[{\includegraphics[width=1in,height=1.25in,clip,keepaspectratio]{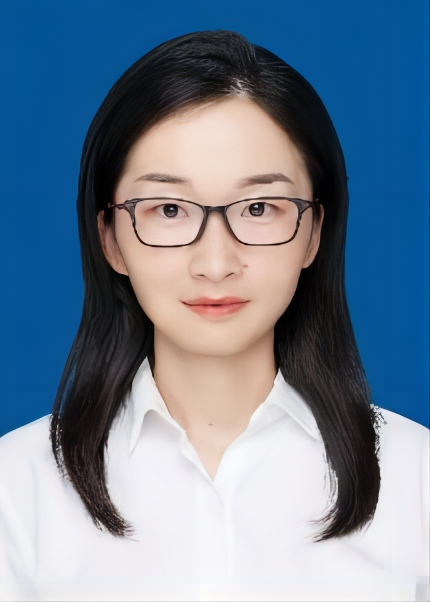}}]{Li Ni}
	received the PhD. degree from University of Science and Technology of China in 2020, and BE degree from Anhui University in 2015. 
	She is presently as a lecturer of School of Computer Science and Technology, Anhui University, Hefei, China. Her research interests include machine learning and data mining.
\end{IEEEbiography}
\vspace{-1.15 cm}
\begin{IEEEbiography}[{\includegraphics[width=1in,height=1.25in,clip,keepaspectratio]{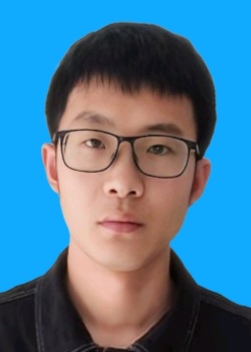}}]{Haowen Shen}
    received the B.S. degree from Anhui University, Hefei, China, in 2023.
    Currently, he is working toward the MSc degree in the School of Computer Science and Technology, Anhui University, China.
    His research interests include machine learning and data mining.
\end{IEEEbiography}
\vspace{-1.15 cm}
\begin{IEEEbiography}[{\includegraphics[width=1in,height=1.25in,clip,keepaspectratio]{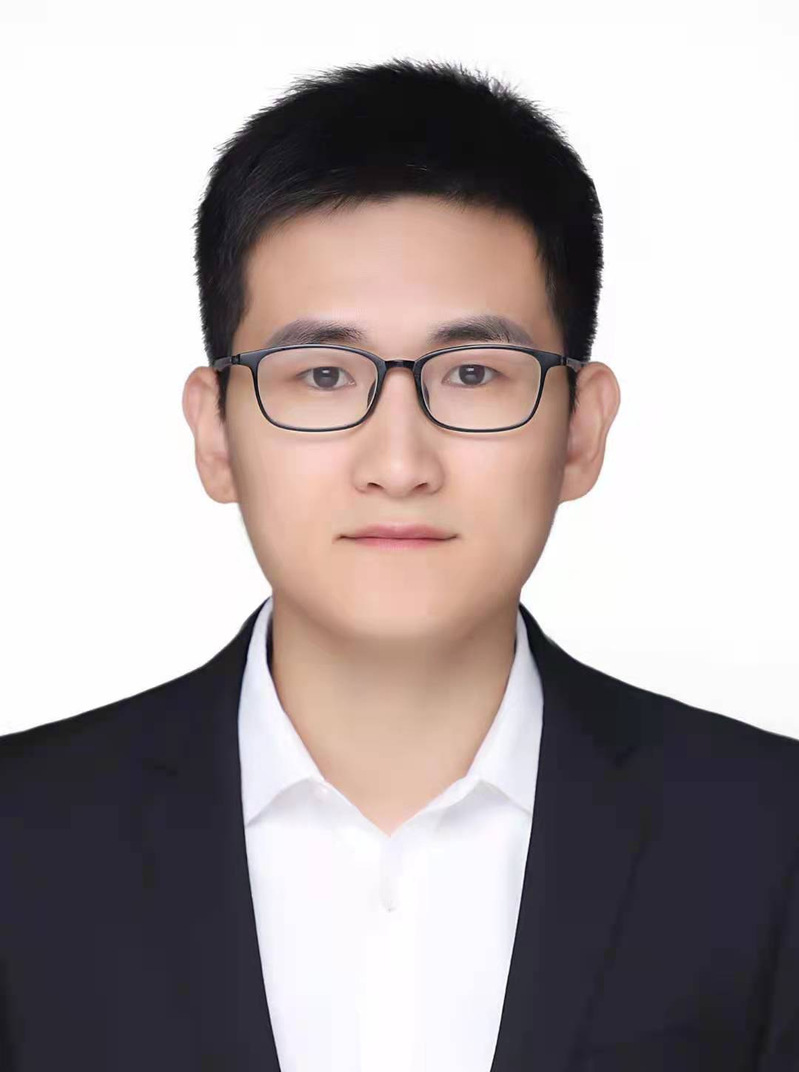}}]{Lin Mu}received the PhD. degree from the University of Science and Technology of China in 2021. He is a lecturer at the School of Computer Science and Technology, Anhui University, Hefei, China. His research interests include information extraction, natural language processing, and large language models (LLMs).

\end{IEEEbiography}
\vspace{-1.15 cm}
\begin{IEEEbiography}[{\includegraphics[width=1in,height=1.25in,clip,keepaspectratio]{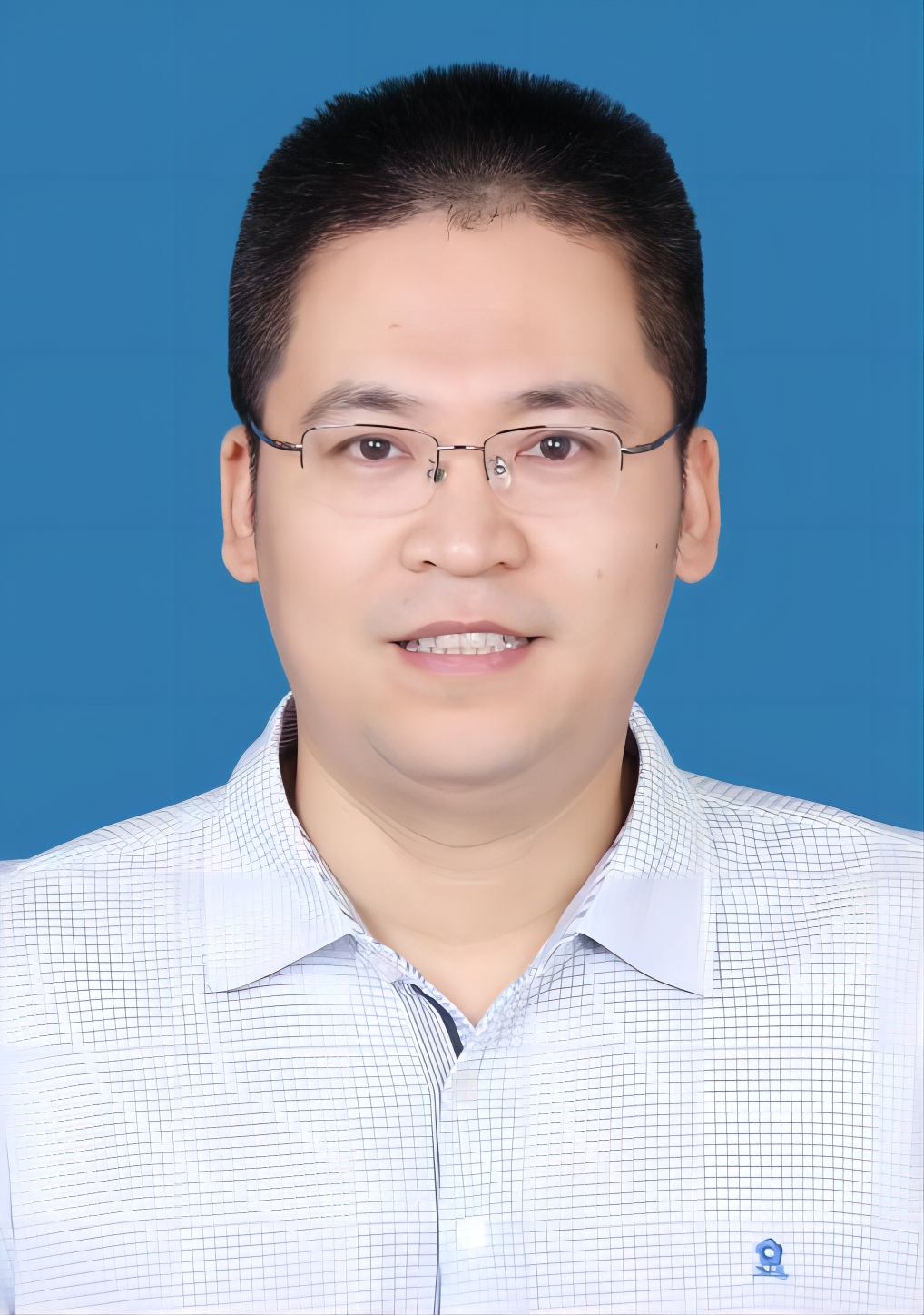}}]
{Yiwen Zhang}
received the Ph.D. degree in management science and engineering from the Hefei University of Technology, Anhui, China, in 2013.
He is a Professor with the School of Computer Science and Technology, Anhui University, China.
His current research interests include service computing, cloud computing, and big data.
\end{IEEEbiography}
\vspace{-1.15 cm}
\begin{IEEEbiography}[{\includegraphics[width=1in,height=1.25in,clip,keepaspectratio]{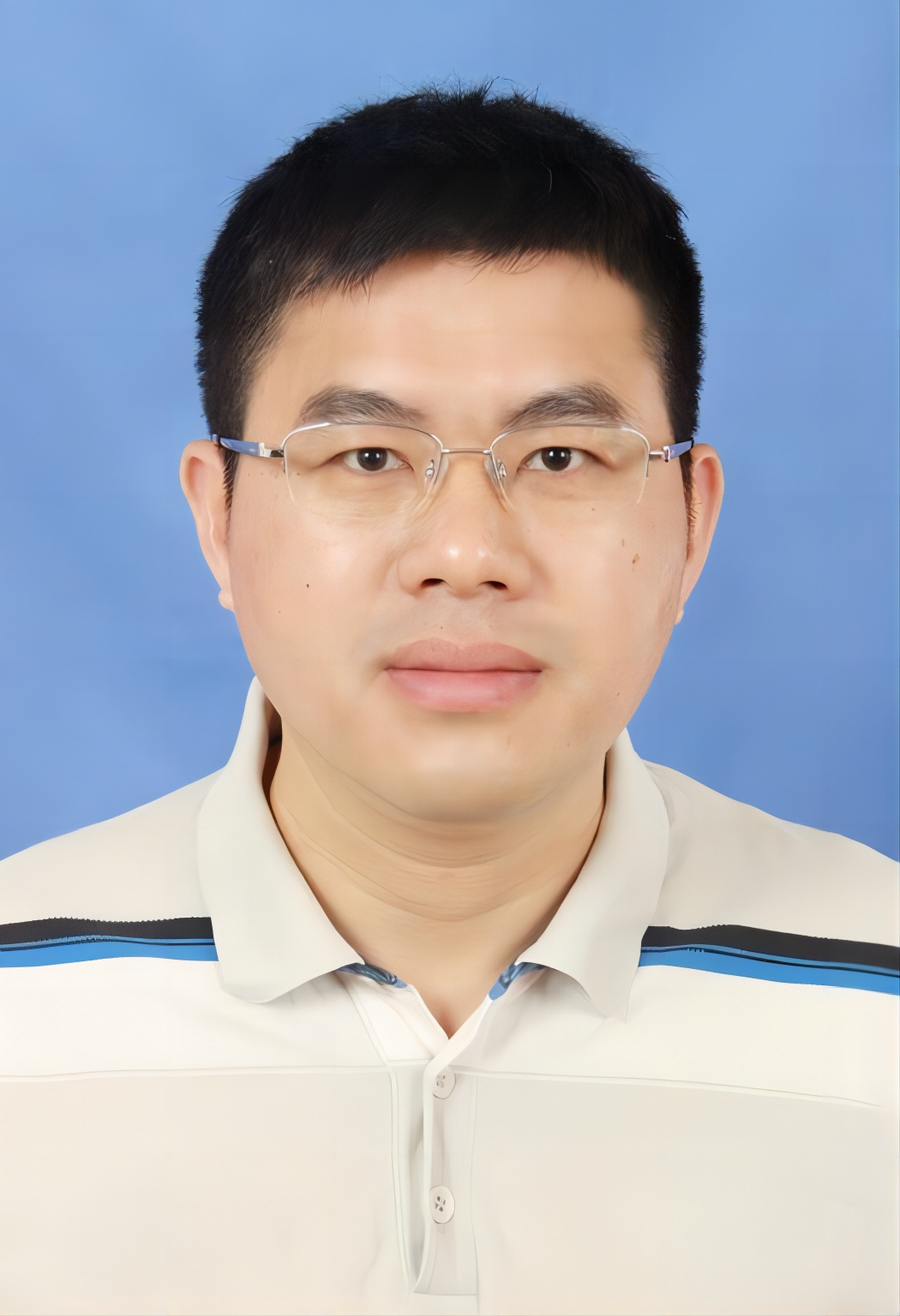}}]{Wenjian Luo}
		received the BS and PhD degrees from Department of Computer Science and Technology, University of Science and Technology of China, Hefei, China, in 1998 and 2003. He is presently a professor of School of Computer Science and Technology, Harbin Institute of Technology, Shenzhen, China. His current research interests include computational intelligence and applications, network security and data privacy, machine learning and data mining. His research interest is artificial intelligence and applications. He mainly focuses on the secure intelligent systems, especially the fundamental algorithms and techniques of machine learning, immune computing, swarm intelligence, and the applications of artificial intelligence in security and privacy, data mining, dynamic and multi-objective optimization. He has published more than 100 international journal and conference papers. He is a distinguished member of CCF and a senior member of IEEE, ACM and CAAI. He currently serves as an associate editor or editorial board member for several journals including Information Sciences Journal, Swarm and Evolutionary Computation Journal, Journal of Information Security and Applications, Applied Soft Computing Journal and Complex \& Intelligent Systems Journal. Currently he also serves as the chair of the IEEE CIS ECTC Task Force on Artificial Immune Systems. He has been a member of the organizational team of more than ten academic conferences, in various functions, such as program chair, symposium chair and publicity chair.
\end{IEEEbiography}

\vfill

\end{document}